\documentstyle[aps,multicol,epsf]{revtex}

\draft
\def\squote{}
\def\quote#1#2#3#4{\squote {#1,\ {\sl#2}\ {\bf#3}, #4}.\par}

\def\prl{{\sl Phys. Rev. Lett.}\ }

\def\pr {{\sl Phys. Rev.}\ }

\def\lr{\ell r}
\def\sss{\scriptscriptstyle}
\def\w{\omega}
\def\e{\epsilon}
\def\deriv{\partial}
\def\A{\sss A}

\def\KS{{\sss KS}}
\def\B{\sss B}
\begin{document}
%\tighten
\title{Van der Waals Energies in  Density Functional Theory}
%}}
\bigskip
\author{\large Walter Kohn$^1$, Yigal Meir$^{2,3}$ and Dmitrii E. Makarov$^{4}$}
\address{$^1$ Department of Physics, University of California at Santa Barbara, Santa Barbara, CA 93106\\
$^2$ Physics Department, Ben Gurion University, Beer Sheva 84105, ISRAEL\\
$^3$ Institute of Theoretical Physics, University of California at Santa Barbara, Santa Barbara, CA 93106\\
$^4$ Department of Chemistry, University of California at Santa Barbara, Santa Barbara, CA 93106}
%\author{\large Yigal Meir}
%\address{Physics Department, Ben Gurion University, Beer Sheva 84105, ISRAEL}
%\author{\large Boris L. Altshuler}
%\address{NEC Research Institute, 4 Independence Way, Princeton, NJ 08540}
\maketitle
\begin{abstract}
In principle, density functional theory yields the correct ground-state
 densities and energies
of electronic systems under the action of a static external potential.
 However, traditional 
approximations fail to include Van der Waals
energies between separated systems.
This paper proposes a practical procedure for remedying this difficulty.
Our method allows seamless calculations between small and large inter-system 
distances. 
The asymptotic H-He and He--He interactions are calculated as a first illustration,
with very accurate results.
\end{abstract}
%\newpage
\begin{multicols}{2}
Density functional theory (DFT)\cite{KS}
 has become a useful tool for calculating
ground-state energies and density distributions of atoms, molecules and
solids, particularly of systems consisting of {\sl many} atoms \cite{review}. 
The simplest 
approximation for practical purposes is the local-density approximation
(LDA) \cite{HK}, based on the properties of the uniform electron gas.
The so-called generalized gradient approximations (GGA) are important
 refinements of
the LDA.

In principle DFT yields the exact ground state energy, including 
long-range Van der Waals (VdW) energies, very important in organic chemistry
and elsewhere. However the commonly used
LDA and GGA, designed for non-uniform
electron gases, fail to capture the essence of VdW energies. The latter
reflect correlated motions of electrons due to the Coulomb interactions
between distant, even non-overlapping atoms, molecules and solids.
Thus a new strategy is needed. 

Here we propose a first-principles approach, which contains the following
essential ingredients: 1. The density distribution, $n(r)$, is approximated
by the LDA or GGA. 2. The Coulomb interaction is divided into short and
long range parts, of which only the latter contributes to VdW energies.
3. The contribution of the long-range interactions  to the energy is expressed
by the so-called adiabatic connection formula (see Eq.(\ref{echi}) below).
4. This expression is transformed into the time-domain, avoiding the need 
to solve a self-consistent equation for the density-density response function.
 As an 
illustration we calculate the asymptotic VdW interaction between two
Helium atoms and between Hydrogen and Helium atoms, with excellent results.
 The method allows seamless calculation of the
interaction of two subsystems, e.g., an atom and a surface, from small to
large separations. Our work was carried out independently and 
differs substantially from recently published work by Anderssen et al
 \cite{anderssen} and by Hult et al.\cite{hult}, which depend critically on a fitting
 parameter.

Since the VdW energies are due to the long range of the
electron-electron interaction, $U(r)=1/r$, we separate this
interaction, as a preliminary step, into short and long range parts, 
\begin{equation}
U(r)=U_{sr}(r)+U_{\lr}(r) .
\label{ul}
\end{equation}
 For example, we can choose  
$U_{sr} (r)\equiv e^{-\kappa r}/r$, with 
$\kappa^{-1}$ chosen somewhat larger than a  typical
intra-atomic electron-electron distance, so that the effect of $U_{\lr}$
on the total energy is small. The calculated total energy is, in principle, 
independent
of the choice of $\kappa$, in practice --- with appropriate approximations --- 
nearly so.
% In what follows $\kappa$ will be a fixed parameter.

We now write the 
Hamiltonian as a function of a coupling constant, ${\lambda}$, that
``turns on" $U_{\lr}$, such that the
physical Hamiltonian operator corresponds to ${\lambda} = {1}$:
\begin{equation}
{\cal H} ({\lambda}) = {T} + {V_\lambda} + {U_{sr}} + {\lambda}{U_{\ell
r}},\hskip .40truein {0}{\leq}{\lambda}{\leq}{1} 
\label{Hlambda}
\end{equation}
where ${T}$ is the kinetic energy, and the external  
potential
${V_\lambda}$ is chosen such that the ground state density
${n_\lambda}({r})$ of ${\cal H}({\lambda})$ equals the exact physical density  
${n}_{\lambda=1}({r})$
for all $\lambda$ \cite{langreth}.
Note that for ${\lambda} = {0}$, the interaction is
entirely short range and that for ${\lambda} = {1}$,  
${V_{\lambda=1}} =
{V_{\rm ext}}$. We denote
 the ground state energy of ${\cal H} ({\lambda})$ by ${E}({\lambda})$.
Then the ground-state energy of the physical system, $E\equiv E(1)$, is given by

\begin{eqnarray}
\label{eint}
E &=& E(0) + \int dr\,\left[ V_{\rm ext}(r)-V_0(r)\right]\,
 n(r) \\
&+& {1\over2} \int dr d r^{\prime}\,U_{\lr}(r-r^{\prime})\left[
\int_0^1 \langle{\hat n}(r) {\hat n}(r^{\prime})\rangle_\lambda \,d\lambda
-n(r)\delta(r-r')\right],\nonumber
\end{eqnarray} 
where ${\hat n}$ is the density operator; $V_0(r)$, defined above, eventually
drops out; see Eq.(\ref{elambda}).

From DFT, $E(0)$ is given by
\begin{eqnarray}
{E} ({0}) &=& T_s [{n} ({r})] + {\int}\,{dr}\,{V_0} ({r}) {n}  
({r})
\nonumber\\
 &+& {1\over2}\,\,{\int}{dr}{d}{r^{\prime}}\,{U_{sr}}  
({r}-{r^{\prime}})
{n}({r}){n}({r^{\prime}}) + {E_{xc}^{sr}} [{n}
({r})]  ,
\label{e0}
\end{eqnarray}
\noindent 
where ${T_s} [{n} ({r})]$, is the non-interacting kinetic energy
functional and 
${E_{xc}^{sr}} [{n} ({r})]$ is the
exchange correlation energy of an electron gas with density ${n}({r})$
and the short range interaction ${U_{sr}}$. 

Substituting (\ref{e0}) in (\ref{eint}) we find, after simple manipulations,
the exact result (independent of the form of $U_{\lr}$):
\begin{eqnarray}			
{E} =  {T_s} [{n} ({r})] &+& {\int}\,{dr}\,
{V_{\rm ext}} ({r})\,{n} ({r})\nonumber\\
&+& \,{1\over2}{\int}\,{dr}{d}{r^{\prime}}
U({r}-{r^{\prime}})\,n({r})n({r^{\prime}})\nonumber\\
&+&{E_{xc}^{sr}} [{n} ({r})]-U_{\lr}(0)N+E_{pol}[{n} ({r})]
\label{elambda}
\end{eqnarray}
where $N$ is the number of electrons, and
\begin{eqnarray}
E_{pol}  \equiv  
{1\over2}\,&\int&\,dr\,dr^{\prime}\,U_{\lr}(r-r^{\prime})\nonumber\\
&\times&\int^1_0d\lambda\, 
\langle({\hat n}(r)-n(r))({\hat n}(r^{\prime})-n(r^{\prime}))\rangle_\lambda
\ . 
\label{epol}
\end{eqnarray}
\noindent $E_{pol}$ includes the
 long-range polarization energies.

To calculate the first four terms in Eq.(\ref{elambda}) we use traditional
methods. Experience \cite{review} has shown that the 
density $n(r)$ may be calculated to a very good approximation
by the LDA
with the full $U(r)$. Such a calculation also automatically yields an 
approximate  $T_s[n(r)]$,
\begin{equation}
{T_s} [{n}] = \displaystyle{\sum^N_{j=1}}\,\,{\epsilon_j} {-} {\int}
{v_{\sss KS}} ({r}) {n} ({r}) \,{dr}, 
\label{ts}
\end{equation} 
\noindent where ${v_{KS}}$ is the Kohn-Sham (KS) effective potential that
reproduces $n(r)$, and 
the ${\epsilon_j}$ are the single-particle energies available from the 
LDA calculation. For $E_{xc}^{sr} [{n} ({r})]$
there exist unpublished excellent results in the LDA \cite{takada}.

To evaluate $E_{pol}$ we use the appropriate exact connection formula 
\cite{langreth}, 
\begin{equation}
E_{pol} = -\int dr dr^{\prime}U_{\lr}(r-r^{\prime})
\int^1_0d\lambda\int^\infty_0 {{d\omega}\over{4\pi}}
{\rm Im} \chi (r,r^{\prime};\omega,\lambda) ,
\label{echi}
\end{equation}
where the linear-response susceptibility  
${\chi }$ is defined, as usual, as follows.
Let $\alpha V_1(r,\w)e^{-i\w t}$ be a small perturbing potential, acting on the
ground state of ${\cal H}(\lambda)$, and producing a density response 
$\alpha\, n_1 (r,\w)e^{-i\w t}$, where $\alpha$ is infitesimal. 
Then  ${\chi }$ is defined by
\begin{equation}
 n_1 (r,\w) = 
\int dr'\,\chi (r,r^{\prime};\omega,\lambda)\, V_1(r',\w)\ .
\label{chi}
\end{equation}

For $\lambda=1$, methods to evaluate
$\chi (r,{r^{\prime}};{\omega},\lambda)$ have been discussed in  
the past \cite{zangwill,gross}, and can be formally carried over
to $\lambda<1$. $\chi$ is the solution of the integral  equation:
\begin{eqnarray}
&\chi&(r,r';\w,\lambda) =  \chi_{\KS}(r,r';\w) + \int dr'' dr''' 
\chi_{\KS}(r,r'';\w)  \nonumber\\
&\times&\left[ U(r''-r''') + f_{xc}(r'',r''';\w,\lambda)\right] 
\chi(r''',r';\w,\lambda) ,
\label{inteqn}
\end{eqnarray}
where $\chi_{\KS}(r,r';\w)$ is the response function of the corresponding
non-interacting Kohn-Sham system, and $f_{xc}$ describes exchange and
correlation effects (see Eq.(6) of Ref.\cite{gross}).

However, except for systems of very high symmetry, such as spherical atoms,
the self-consistent solution of (\ref{inteqn}) is computationally very
forbidding. Here we propose an equivalent but much less cumbersome 
procedure, which avoids the solution of a self-consistent integral
equation for each value of $\lambda$. We note that  $\chi(r,r';\w,\lambda)$
 is the
Fourier transform, $\chi(r,r';\w,\lambda)=\int dt\, \chi(r,r';t,\lambda) 
e^{i\w t}$, of 
the {\sl time-dependent} response function, $\chi(r,r';t,\lambda)$, 
defined as follows:
\begin{equation}
n_1 (r,t,\lambda) = 
\int dr' dt' \,\chi (r,r^{\prime};t-t',\lambda)\, V_1(r',t')\ \  ,
\label{chit}
\end{equation}
where $ V_1(r',t)$ and $n_1 (r,t,\lambda)$ are, respectively, external 
perturbing potential
and density response. Eq.(\ref{echi}) can be rewritten as 
\begin{equation}
E_{pol} = -{1\over{4 \pi}} \,\int\, dr\, dr^{\prime}\, U_{\lr}(r-r^{\prime})\,
\int^1_0\,d\lambda\,\int^\infty_0\, {{dt }\over{t }}\,
\,\chi (r,r^{\prime};t,\lambda) .
\label{echit}
\end{equation}
Following Gross and Kohn \cite{gross}, we can replace the density response
of the 
physical system to the external perturbing potential by the response of the 
($\lambda-$
independent) KS system to an effective potential,
\begin{equation}
n_1 (r,t,\lambda) = 
\int dr' dt' \,\chi_{\KS} (r,r^{\prime};t-t',\lambda) 
\,V_1^{eff}(r',t',\lambda)\ \  ,
\label{chiks}
\end{equation}
where
\begin{eqnarray}
&V&^{eff}_1(r',t',\lambda) =  V_1(r',t') + 
 V_{1,xc}(r',t',\lambda) \nonumber\\&+& 
 \int dr'' \left[U_{sr}(r'-r'') + 
\lambda U_{\lr}(r'-r'') \right] n_1(r'',t,\lambda)\  .
\label{veff}
\end{eqnarray}
\noindent  $V_{1,xc}(r',t',\lambda)$ is defined by Eqs.(\ref{chiks}) and
 (\ref{veff}).
Thus $\chi(r,r';t,\lambda)$ is the density response of the non-interacting
 KS system at
time $t$, to the $ V^{eff}_1(r',t',\lambda)$ at $0~\le~t'~\le~t$, induced by
 an external
potential $V_1(r'',t')=\delta(r''~-~r)\delta(t)$.

To complete this procedure we need a practical approximation for $V_{1,xc}$,
 for
which, following Ref,\cite{zangwill}, we set
\begin{equation}
V_{1,xc}(r',t',\lambda) = \left.  {{\deriv V_{xc}(n,\lambda)}\over{\deriv n}}
\right|_{n_0(r')} 
n_1(r',t',\lambda)\ ,
\label{v1xc}
\end{equation}
where $n_0$ is the unperturbed density and $V_{xc}$ is the static 
exchange-correlation
potential in the LDA. Here, in addition to the usual approximation of the
 LDA, 
the frequency
dependence (or retardation) of  $ V_{xc}$ is neglected.

The evaluation of $\chi$ now requires the calculation of the evolution of the 
non-interacting
KS system under the action of $V_1^{eff}(r',t',\lambda)$. At this point it is
 convenient to
change from the coordinate representation to an orthonormal basis, $f_n(r)$,
 and write
generically for any $F(r)$ and $G(r,r')$
\begin{eqnarray}
F(r) &=& \sum_{m=1}^\infty F_m f_m(r)\nonumber\\
 G(r,r') &=&  \sum_{m,m'=1}^\infty
G_{m\, m'}\, f_m(r) f_{m'}(r') .
\label{basis}
\end{eqnarray}
Thus Eq.(\ref{chit}) becomes
\begin{equation}
n_{1,m}(t,\lambda) = \int_0^{\infty} dt' \sum_{m'} \chi_{\KS,m\,m'}(t-t';\lambda)
\,V_{1,m'}^{eff}(t',\lambda) .
\label{chibasis}
\end{equation}
The following steps need to be carried out for each value of $\lambda$:  1. At 
time 
$t=0^{\sss -}$
the KS system is given by the determinant  $(N\!)^{-1}{\rm Det}\left| 
\phi_1\,\phi_2\ldots 
\phi_N\right|$ of the occupied KS orbitals $\phi_j$. 2. At time $0^{\sss +}$, 
after the
 action of a
small external perturbation, $V(r)=\alpha f_m(r) \delta(t) $ ($\alpha$ small),
 each of the KS
orbitals is changed, $\phi_j(r,t)\rightarrow \phi_j(r,t) - i\alpha f_m(r) \phi_j(r,t)$. 
(Effects on the orbitals of the finite unperturbed KS Hamiltonian and of
the induced parts of $V^{eff}$ in the infinitesimal interval $(0^-,0^+)$ are 
negligible).
3. For $t>0^{\sss +}$ we integrate the time-dependent Schr\"odinger equation
 for each 
$\phi_j$ in a stepwise fashion, evaluating 
the first-order induced density $\alpha n_{1,m}(r,t) = \sum_j  
|\phi_j(r,t)|^2 - n_0(r)$ at each time step, to be able to compute the induced 
parts of
$V_1^{eff}$ (Eq.(\ref{veff})), which depend on $n_{1,m}(r,t)$. 4. The projection  of 
$n_{1,m}$ on $f_{m'}$ gives $\chi_{m\,m'}(t)$,
\begin{equation}
\chi_{m\,m'}(t) = \left( f_{m'}, n_{1,m}(t) \right).
\label{chimm}
\end{equation}
\noindent  From Eq.(\ref{echit}), we obtain
\begin{equation}
E_{pol} = - {1\over 4\pi} \int_0^\infty {dt\over t} \int_0^1 d\lambda 
\sum_{m,m'} \chi_{m\, m'}(t,\lambda)\, U_{\lr;m\, m'} .
\label{echimm}
\end{equation}
In practice, the integration over $\lambda$ is replaced by a finite sum.

As a simple example of this general procedure, we now apply it to the 
calculation of the asymptotic VdW  interaction of a pair of spherically
symmetric atoms. We denote the atoms by $A$ and $B$, and their nuclear
coordinates by $R_A$ and $R_B$
 (taken to be on the $z$-axis), and write $R=|R_A-R_B|$.
We take $R\gg a_A+a_B$, the sum of the atom radii and $\kappa \simeq
 (R a)^{\sss -1/2}$,
where $a$ is the atomic  radius. The asymptotic VdW interaction is obtained 
from those
parts of $E_{pol}$ (Eq.(\ref{epol})) in which $r$ and $r'$ are in different 
atoms. Take
$r$ to be in $A$ and $r'$ in $B$. Next we write $U=U_{sr}+\lambda U_{\lr}$. 
Since $\kappa\rightarrow 0$ when $R\rightarrow \infty$, $ U_{\lr}$
can be treated as a small perturbation, giving to first order
\begin{eqnarray}
\chi(r,r^{\prime};\omega,\lambda) &=& 
\lambda \int \,dr_1 \,dr_2\, U_{\lr}(r_1-r_2)\nonumber\\
&\times&\chi_{\A}(r,r_1;\omega)\, \chi_{\B}(r_2,r^{\prime};\omega)   ,
\label{chiab}
\end{eqnarray}
where $\chi_{\A} (\chi_{\B})$ is the response of the isolated atom $A\,(B)$.
The integration over $\lambda$ is now trivial. 
Lastly, we expand $U_{\lr}(r-r^{\prime})$ in $1/R$ and obtain the final 
expression, $E_{VdW}=-C_6/R^6$,
\begin{eqnarray}
C_6 &=& {3\over{\pi}}\,\, {\rm Im} \int_0^\infty 
d\w\,\chi_{\A}^{zz}(\w) \,\chi_{\B}^{zz}(\w) \nonumber\\
&=&  {3\over{\pi}} \int_0^\infty dt_1 \int_0^\infty dt_2\,
 {{\chi_{\A}^{zz}(t_1)\,
\chi_{\B}^{zz}(t_2)}\over{t_1+t_2}} .
\label{eVdW}
\end{eqnarray}
In the above $\chi^{zz}$ is defined as the $z$-component of the density
response to a perturbation in the $z$-direction,
 $\chi^{zz} = \int\, dr_1 \,dr_2 \,\chi(r_1,r_2)\,z_1 \,z_2$.
The first form is well known,  the second its Fourier transform into the time
domain.

We have calculated the time-dependent response for the Helium atom in DFT
as follows. We begin with the exact $V_{xc}(r)$ \cite{umrigar}, which reproduces the 
exact ground-state density $n_0(r)$ (known from highly accurate independent calculations),
and the corresponding exact KS ground-state wavefunction $\phi_0$ and energy $\e_0$.
We take as
 perturbation $V_1(r,t)=-\alpha z \delta(t)$.
  At time $t=0^{\sss +}$ the wavefunction will be
\begin{equation}
\phi(r,t=0^{\sss +}) = \phi_0(r) - i\alpha z \phi_0(r)\ \ ,
\label{phi0}
\end{equation}
a combination of s- and p-like functions. For $t>0$, we solve the time-dependent
Schr\"odinger equation for $\phi(r,t)\equiv \phi_0(r,t) + \alpha \phi_1(r,t)$, 
with the initial condition (\ref{phi0}). Linearizing in $\alpha$
gives the following equation for $\phi_1$:
\begin{eqnarray}
{i}{{\deriv\phi_1(r,t)}\over{\deriv t}} &= &{\cal H}_0 \phi_1(r,t) + 
{\cal H}_1(t) \phi_0(r,t)
 \ , \nonumber \\
\phi_1(r,0^{\sss +})& = &{-i}r \phi_0(r)
\label{phi1}
\end{eqnarray}
where $\phi_0(r,t) = e^{-i\e_0t}\phi_0(r)$, ${\cal H}_0$ is the KS 
unperturbed Helium Hamiltonian,
\begin{equation}
{\cal H}_1(t) = \int \,dr'\,{{n_1(r',t)}\over{|r-r'|}} \ + \  \left.
{{\deriv V_{xc}}\over{\deriv n}} \right|_{n_0(r)} n_1(r,t) ,
\label{h1}
\end{equation}
and  $n_1(r,t)=4 Re\left[\phi_0(r,t)\phi_1^*(r,t)\right]$. 
%Defining $R_i(r)=r  \phi_i(r)$, the equation for $R_1$ takes the form:
%\begin{eqnarray}
%i{{\deriv R_1(\r)}\over{\deriv t}} = \left[ -{1\over2}{{\deriv^2}\over{r^2}} + 
%{1\over {r^2}}\right] R_1(\r) + V_0^{eff}(\r)R_1(\r) &+& {{4\pi}\over3} \int_0^\infty
% 2{\rm Re}\left[ R_0(\r')R_1^*(\r')\right]{{r_<}\over{r_>^2}} d\r' R_0(\r) \noindent\\
%&+& \  \left.{{\deriv V_{0,xc}}\over{\deriv n}} \right|_{n_0(\r)}
%2{\rm Re}\left[ R_0(\r')R_1^*(\r')\right]R_0(\r)/\r^2
%\label{r1}
%\end{eqnarray}
% where the $t$-dependence of $R_0$ and $R_1$ has been 
$V_{xc}$ was calculated using the parameterization of Vosko et al. \cite{vosko}.
This equation was solved by stepwise integration in time.
The time evolution from $t$ to $t+\Delta t$ was carried out using the fast Fourier 
transform method as used in Ref.\cite{feit}.
 Since at each instant 
$\phi_1$
evolves under the action of the total effective potential, the resulting
response function
$\chi(t)$ (and, if desired, the corresponding $\chi(\w)$) is automatically 
self-consistent 
{\sl without the
need to first solve a self-consistent integral equation}, as is the case
 in the direct evaluation
of $\chi(\w)$ (see Eq.(\ref{inteqn})).

In practice, the direct evaluation of the time integral in (\ref{eVdW}) is
inconvenient because $\chi(t)$ oscillates with undiminishing amplitude at large $t$. 
We have therefore noted that if we define 
$\alpha(u)=\int_0^\infty \chi^{zz}(t) e^{-ut} dt$ 
(i.e. $\alpha(u)=\chi^{zz}(i\w)$), the
VdW coefficient, $C_6$, can be written as
\begin{equation}
C_6 = {3\over{\pi}}\,\,  \int_0^\infty 
du\,\alpha_{\A}(u) \,\alpha_{\B}(u) .
\label{c6}
\end{equation}

For helium $\chi(t)$ was calculated up to $t=15$ atomic units (AU), which allows 
accurate calculation of $\alpha(u)$ for $u>u_0=0.4$. In the interval
$0\le u\le u_0$, we represented $\alpha(u)$ by the expression $a+b/(1+cu^2)$, and 
fitted $a,b,c$ to $\alpha(u)$ and its first two derivatives at $u=u_0$. (We checked
that the results are insensitive to the exact choice of $u_0$ or to thw choice of the
extrapolating function). Fig.~1 shows our 
$\alpha(u)$ for He. The correct asymoptotic form, $\alpha(u)\rightarrow 2/u^2$ (the
f-sum rule), is automatically obeyed. The completeness sum rule, requires
$\int_0^\infty \alpha(u) du = 2\pi<\phi_0|z^2|\phi_0> \simeq  2.50$. Our $\alpha(u)$
gives $2.33$. An independent check on our $\alpha(u)$ is the static susceptibility
$\alpha(0)$. The best theoretical value is  $1.383241$ \cite{susceptibility}, 
while we find $1.38$.

\begin{center}
\leavevmode
\epsfxsize=3in
\epsfbox{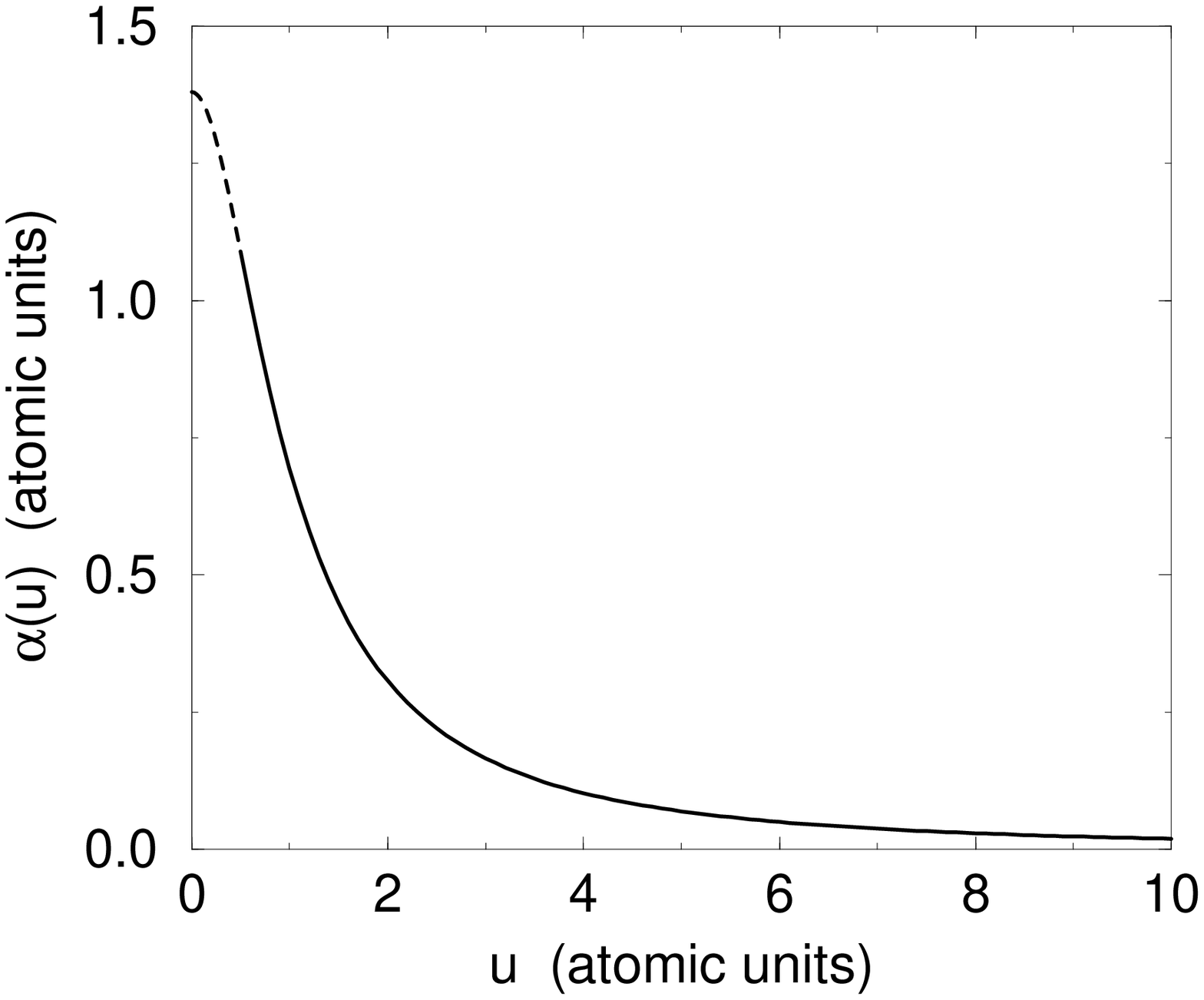}
%\epsfbox{qot.eps}
\end{center}
\begin{small}
\centerline{Fig. 1. The imaginary-frequency susceptibility $\alpha(u)$ for Helium}
\centerline{(solid line - direct evaluation, broken line - extrapolation).}
\end{small}
\vskip 0.5 truecm

Our results for the He-He VdW constant is $C_6=1.45$, almost identical to the best
theoretical value \cite{HeHe} $1.458$. For the H-He system we find $C_6=2.81$ 
compared to the best theoretical value \cite{HHe} of $2.817$.

We feel cautious about the significance of the high accuracy of our results for
the He-He and the H-He systems in view of the fact that our calculated $\alpha(u)$
leads to a $7\%$ error in the completeness sum rule. At the same time our results
demonstrate the soundness and feasibility of our approach. We are optimistic that
our approach will not only give asymptotic van der Waals coefficients, but the 
{\sl entire} nuclear 
potential energy function $\epsilon(R)$, including polarization energies.

We found that the results are rather sensitive to the choice of a good KS
potential for the unperturbed ground state. Repeating the calculation by
replacing the exact $V_{xc}$ by $V_{xc}$ in the local density approximation,
the result for $C_6$ of the He-He system was $1.85$, $28\%$ too high.  This is
qualitatively similar to the experience of Petersilka et al.\cite{petersilka}
with calculations of excited-state energies.

We are indebted to C. Umrigar for providing us with the exact KS and the LDA KS 
data for Helium.
We also thank Weitao Yang for helpful information. This project was 
supported by the US National Science Foundation, Grant No. DMR-9630452, and
grant No. PHY94-07194, and by 
the US-Israeli
Binational Science Foundation, Grant No. 94-00277/1.

\end{multicols}
\end{document}